\documentclass{article} 
\usepackage{arxiv_conference,times}

\usepackage{amsmath,amsfonts,bm}

\def\eqref#1{equation~\ref{#1}}

\def\1{\bm{1}}

\DeclareMathAlphabet{\mathsfit}{\encodingdefault}{\sfdefault}{m}{sl}
\SetMathAlphabet{\mathsfit}{bold}{\encodingdefault}{\sfdefault}{bx}{n}

\usepackage{hyperref}
\usepackage{url}

\usepackage{graphicx}  
\usepackage{booktabs}
\usepackage{multirow}

\usepackage{caption}
\usepackage{wrapfig}

\usepackage[table]{xcolor}
\definecolor{bestgreen}{RGB}{169,208,142}   
\definecolor{secondgreen}{RGB}{226,239,218} 

\newcommand{\modelicon}[2][0 0 0 0]{%
  \makebox[1.55em][c]{%
    \raisebox{-0.18\height}{%
      \includegraphics[
        height=1.05em,
        trim=#1,
        clip,
        keepaspectratio
      ]{#2}%
    }%
  }%
  \hspace{0.18em}%
}

\newcommand{\modelheader}{%
  \hspace*{1.73em}{\large\textbf{Model}}%
}

\title{Code Skill Server: A Large-Scale Search Engine for Code Generation Skills}

\author{
\parbox{0.98\textwidth}{
\centering
\textbf{Xiaoyong Li}$^{1,*}$,
\textbf{Haolong Li}$^{1,*}$,
\textbf{Dongliang Xu}$^{1,\dagger}$,
\textbf{Haoyu Li}$^{1}$,
\textbf{Yifan Wang}$^{1}$,
\textbf{Jingyuan Shi}$^{1}$,
\textbf{Yanghua Wang}$^{1}$,
\textbf{Yanbin Sun}$^{2}$,
\textbf{Shidong Pan}$^{3}$,
\textbf{Yue Yao}$^{1,\dagger}$
\\[6pt]
\normalfont
$^{1}$Shandong University,
\quad
$^{2}$Guangzhou University,
\quad
$^{3}$Independent Researcher
\\[3pt]
\normalfont
$^{*}$Equal contribution
\qquad
$^{\dagger}$Corresponding authors
}
}

\iclrfinalcopy

\begin{document}

\maketitle

\begin{abstract}

Agent skills have proven to be an effective technique for improving agent ability on specific tasks, including software engineering, natural science, and mathematics. During such a process,  using agent skills often involves manually discovering and organizing skills from heterogeneous sources, while existing skill retrieval methods mainly focus on selecting relevant skills from established skill libraries.
However, with the rapidly increasing number and variety of reusable agent skills, manually searching for relevant  skills across heterogeneous sources is becoming impractical at scale. In this paper, we introduce the Code Skill Server (CSS), a large-scale search engine for finding the most useful code generation skills for a target task. 
CSS consists of a skill server that incrementally collects and indexes publicly available code generation skills from multiple sources, with 993 skills included in the corpus used in this work. Extensive experiments across different retrieval methods and code generation benchmarks show that the best CSS configuration improves Pass@1 by up to 13.91\%  and consistently improves all evaluated models by at least 4.21\% in the cross-model evaluation. 
These results demonstrate that CSS can effectively search for and serve relevant skills, improving code generation quality and enhancing models' code generation capabilities. Code and the CSS-993 corpus are available at
\url{https://github.com/haoyuli-sdu/CodeSkillServer}.
\end{abstract}

\section{Introduction}
\label{sec:introduction}

Large language models have advanced code generation by translating natural language specifications into programs, while code generation agents have further extended its application to tasks within existing software projects \citep{repocod,swebench}. Despite these advances, code generation for practical applications remains challenging: models need to understand complex task requirements, use external libraries appropriately, and account for dependencies in the surrounding project \citep{zhuo2025bigcodebench,classeval}. These challenges have motivated researchers to explore methods that provide code generation agents with additional domain knowledge and procedural guidance to complement the knowledge already encoded in the models.

Agent skills provide a reusable form of task guidance. Voyager demonstrates how an agent can accumulate and retrieve executable skills, while SkillsBench studies structured skill packages consisting of instructions and supporting resources and evaluates their effects on task performance \citep{wang2024voyager,li2026skillsbench,pan2026skillguard}. For code generation, skills can describe  recurring implementation procedures, testing patterns, and architectural conventions \citep{gskill}. However, having a collection of skills does not determine which skills are appropriate for a particular task. Manual selection requires users to search candidate skills across multiple skill sources, inspect their instructions, and assess their applicability. Differences in terminology, scope, and prerequisites across skills make this process more complex than matching a task to a broad capability label. Recent retrieval methods have focused on important parts of this process, such as selecting relevant skills from a skill library and organizing them into usable contexts \citep{skillrouter,liu2026graphofskills,liao2026understanding}. However, these methods mainly focus on retrieval and contextualization within a skill library; a complete workflow also needs to connect the collection and organization of skills from external sources with skill retrieval and subsequent use by code generation agents. Therefore, a framework is still needed to integrate these stages into a unified service.

\begin{wrapfigure}[18]{l}{0.63\textwidth}
    \centering
    \vspace{-8pt}
    \includegraphics[width=\linewidth]{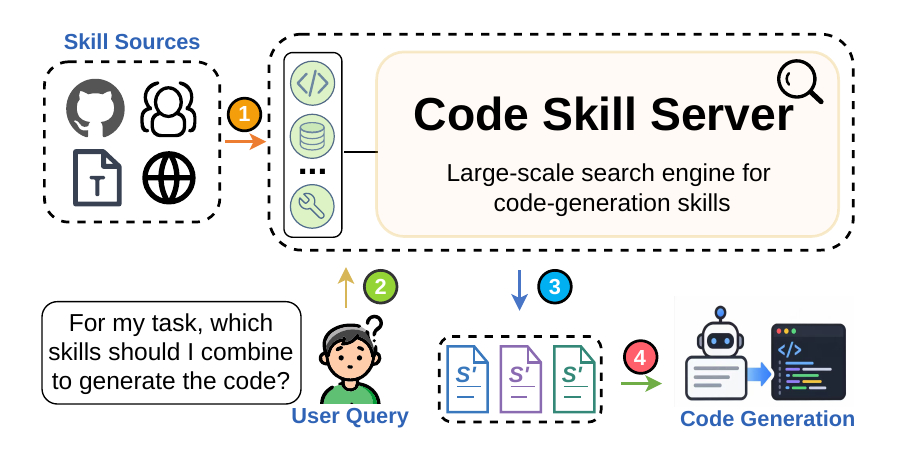}
    \caption{Code Skill Server: A large-scale search engine for identifying and combining relevant code generation skills. The server indexes skills from diverse sources, retrieves relevant skills according to the user query, and provides the selected skills to guide code generation.}
    \label{fig:mov}
    \vspace{0pt}
\end{wrapfigure}

To address this issue, we propose Code Skill Server (CSS), a large-scale search engine for code generation skills that connects skill search, organization, retrieval, and client use into a unified framework, as shown in Fig.~\ref{fig:mov}. CSS consists of a SkillServer and clients: the SkillServer prepares and organizes code generation skills and retrieves relevant skills for target tasks, while clients are equipped with code generation agents that use the retrieved skills. CSS prepares publicly available skills as a shared resource, represents them in a consistent format, organizes them by capability, and builds indexes for retrieval. Using this workflow, we have collected and organized an initial set of 993 code generation skills. When a client receives a code generation task, FASTADAPT uses the client model to construct a structured skill retrieval request, and the server applies the configured retrieval method to return relevant skills to the client. The retrieved skills are provided to the client's code generation agent, which determines how to use them. The server determines which skills are served, while the agent determines which of the served skills to use. This separation enables different retrieval methods and client models to use the same prepared skill collection through a unified request and response interface, without requiring skills to be repeatedly collected and organized for individual tasks.

We evaluate CSS on two well-established code generation benchmarks, BigCodeBench  \citep{zhuo2025bigcodebench} and CoderEval \citep{yu2024codereval}, covering function generation using external libraries and code generation within existing project contexts. We compare three retrieval configurations over a shared corpus of 993 skills. The best-performing CSS configurations improve task pass rates from 53.42\% to 60.88\% on BigCodeBench and from 33.04\% to 46.96\% on CoderEval. We further evaluate CSS across five models, and separately examine the effects of skill collection size and curated versus raw skill collections on task performance. Taken together, the evaluation demonstrates the practical utility of CSS for serving code generation skills to code generation agents across different retrieval methods and client models.

Overall, our contributions are threefold: (1) We introduce CSS, a large-scale search engine that connects code generation skill search, organization, retrieval, and agent use. (2) We construct a curated collection of 993 code generation skills with standardized records and capability labels. (3) We evaluate CSS on two code generation benchmarks and five models, comparing three retrieval methods and analyzing task-level outcomes and the effects of collection size and composition.

\section{Related Work}

\textbf{Agent Skills.} Recent studies improve LLM agent capabilities through reusable skills, experience-derived knowledge, and workflows without updating model parameters \citep{wang2024voyager,zhao2024expel,wang2025awm}. Voyager maintains an ever-growing skill library of executable code for storing and retrieving complex behaviors \citep{wang2024voyager}. ExpeL extracts reusable knowledge from past experiences to guide decisions on new tasks \citep{zhao2024expel}. Agent Workflow Memory induces reusable workflows from agent trajectories to guide future actions \citep{wang2025awm}. More recent work explores automatic skill discovery and evolution. SkillWeaver enables web agents to autonomously discover and refine skills \citep{zheng2025skillweaver}. SkillRL introduces a recursive evolution mechanism that allows its skill library and agent policy to evolve together during reinforcement learning \citep{xia2026skillrl}. As skill libraries scale, efficient skill organization and retrieval become increasingly important. Graph-of-Skills models dependencies among skills to enable dependency-aware retrieval, while SkillTrace combines query decomposition, query-skill matching, and skill dependencies for compositional skill retrieval \citep{liu2026graphofskills,yao2026skilltrace}. SkillsBench provides a systematic benchmark for evaluating skill effectiveness by comparing agent performance with and without skills across diverse tasks \citep{li2026skillsbench}. Complementing these studies, we organize externally collected skills into a shared resource for code generation and provide a common serving interface for comparing retrieval methods.

\textbf{Code Generation.} Code generation is an important application of large language models, motivating benchmarks that evaluate functional correctness from natural-language specifications. HumanEval evaluates function-level code generation through executable unit tests \citep{chen2021codex}. DS-1000 focuses on realistic data science tasks involving external libraries, whereas EvalPlus strengthens functional correctness evaluation through expanded test suites \citep{lai2023ds1000,liu2023evalplus}. Subsequent benchmarks have moved toward more realistic programming settings. CoderEval evaluates functions and methods from real-world open-source projects while considering surrounding project dependencies \citep{yu2024codereval}. BigCodeBench extends function-level evaluation to complex tasks involving real-world libraries and tests whether models can compose multiple function calls to solve practical programming problems \citep{zhuo2025bigcodebench}. These benchmarks primarily assess model capabilities rather than the effects of externally sourced reusable skills \citep{yu2024codereval,zhuo2025bigcodebench}. We use CoderEval and BigCodeBench to evaluate whether retrieving such skills improves the functional correctness of generated code.
\vspace{-5pt}
\section{Code Skill Server}
\label{sec:css}

\begin{figure}[t]
    \centering
    \includegraphics[width=\linewidth]{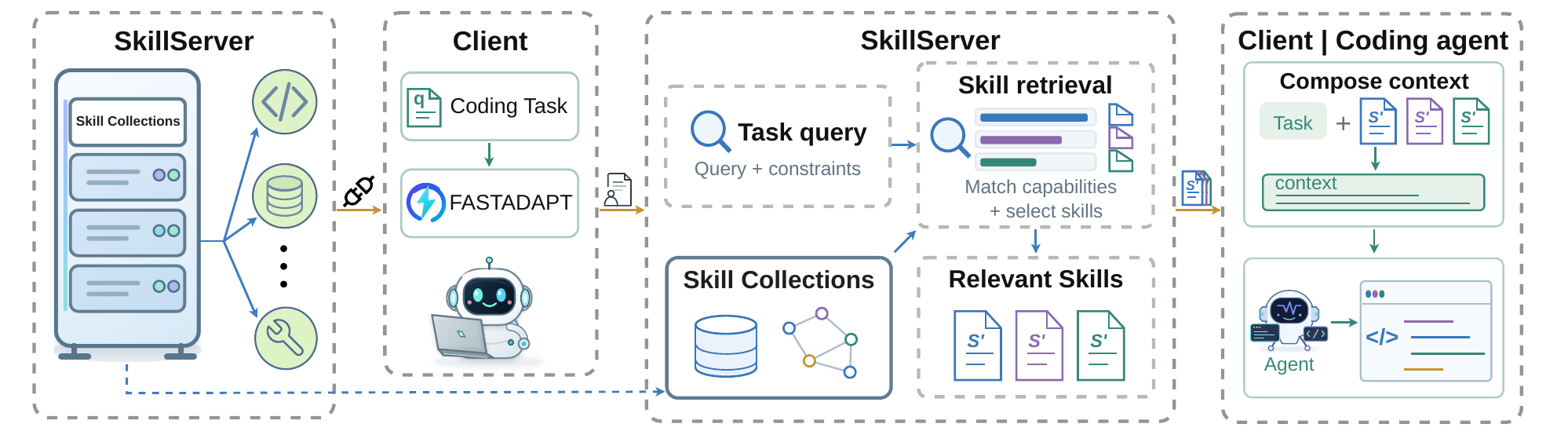}
    \caption{Overview of Code Skill Server (CSS). CSS consists of a SkillServer that maintains an indexed collection of code generation skills from diverse sources and clients equipped with code generation agents. Given a coding task, FASTADAPT constructs a structured skill search request specifying the task query and retrieval constraints. The SkillServer retrieves relevant skills from its  indexed collection and serves them to the client, where the code generation agent determines how to use the retrieved skills to complete the task.}
    \label{fig:overview}
    \vspace{-10pt}

\end{figure}

Code Skill Server (CSS) is a search engine for retrieving relevant skills for code generation tasks. As shown in Fig.~\ref{fig:overview}, CSS consists of a SkillServer that maintains an indexed collection of skills and a client equipped with a code generation agent. Given a target coding task, the client submits a skill search request, the server retrieves relevant skills using the configured retrieval method, and the client agent decides how to use them during code generation.

Skills collected from diverse code repositories and skill registries differ in document structure, metadata, and the capabilities they describe. Names alone are insufficient to identify them: different skills may share a name, while identical skill documents may appear at multiple source locations. CSS acquires and filters the skills, normalizes them into structured records, and organizes them into capability pools. Searchable indexes enable efficient retrieval from the prepared collection. Given a coding task, the client constructs a query with retrieval constraints, the server selects and returns ordered skills, and the agent decides which of the served skills to use during code generation.

\subsection{PROBLEM DEFINITION}
\label{sec:css_problem}

Let $S=\{s_1,\ldots,s_N\}$ denote the collection of code generation skills indexed by the SkillServer, and let $q$ denote a target coding task from a client. A task consists of a natural language specification together with its available coding context, such as the programming language, function signature, library dependencies, and input and output requirements. Each skill provides reusable implementation guidance for a class of coding tasks rather than a reference solution for a specific task instance.

Our goal is to select a subset of skills $S_q\subseteq S$ that is relevant to the target task $q$. Let $S'\subseteq S$ denote a candidate skill subset. The ideal selection is formulated as
\begin{equation}
    S_q^*
    \in
    \operatorname*{arg\,max}_{S'\subseteq S}
    \operatorname{Rel}(q,S'),
    \label{eq:css_objective}
\end{equation}
where $\operatorname{Rel}(q,S')$ measures the relevance of the candidate skill subset $S'$ to the target task $q$. 

Intuitively, given a target coding task, we aim to find the skills that are most relevant to the requirements of the task. The selected skills are then provided to the client coding agent, which determines how to incorporate them into the code generation process.

The problem becomes challenging when the skill collection is large and heterogeneous. Skills collected from different repositories and registries may have different document structures, metadata, naming conventions, and capability descriptions. Moreover, a client should not need to manually inspect the entire collection for every coding task. CSS addresses this problem by searching for and organizing skills on the SkillServer, then retrieving relevant skills based on each client task.

\vspace{-10pt}
\subsection{SERVER DESIGN}
\label{sec:css_server}

The SkillServer prepares the collection independently of client requests. This stage turns heterogeneous external skills into identifiable, searchable resources. We first describe discovery, acquisition, filtering, and normalization, then define the capability pools used to organize the resulting collection.

\subsubsection{SKILL SEARCH AND ORGANIZATION}
\label{sec:css_organization}

\paragraph{Discover skills.}
CSS searches public code repositories and skill registries, including GitHub, SkillsMP, and Skills.sh, for potential code generation skills. Discovery uses queries that describe specific coding tasks and the code they are intended to produce, such as function implementations and application scaffolds. These queries target code-producing capabilities rather than relying only on broad terms such as a programming-language name. Each discovered occurrence retains its available title, description, source location, and query provenance.

\paragraph{Acquire and filter.}
For each candidate, CSS locates and reads the corresponding \texttt{SKILL.md} to obtain the actual skill instructions. Repository and path information are used to identify the intended document and distinguish it from other skills in the same repository. Candidates that cannot be resolved unambiguously are retained for further inspection rather than being discarded. CSS then filters the acquired skills based on whether they provide instructions for producing or transforming code. The decision considers evidence from the skill name, description, metadata, and instruction body, while excluding skills focused only on explanation, planning, review, or other non-code outputs. For each candidate, CSS records the supporting evidence and filtering decision, enabling manual inspection of ambiguous cases.

\paragraph{Normalize skills.}
CSS converts the retained documents into a common record structure containing metadata, instructions, and provenance. Names and descriptions are extracted from the available document fields, while the original instructions and source information are preserved. CSS removes exact duplicates using SHA-256 hashes of the acquired Markdown content, storing identical content as a single record while retaining different source locations as separate provenance records. Documents with the same name but different contents remain distinct.

\paragraph{Build skill indexes.}
The normalized records provide the input for skill retrieval. CSS supports two types of retrievers over the prepared skill collection: content-based retrievers use textual fields for task and skill matching, while relation-based retrievers additionally use an index of skill relations. Both types operate on the same skill records. CSS provides a common interface for different retrieval methods, and the prepared indexes are reused across requests rather than rebuilt for each request.

\subsubsection{Skill Capability Pools}
\label{sec:css_pools}

CSS represents each skill as
\begin{equation}
    s_i=(n_i,d_i,t_i,c_i,p_i),
    \label{eq:css_record}
\end{equation}
where $n_i$ is its name, $d_i$ its description, $t_i$ its instruction text, $c_i$ its primary capability category, and $p_i$ its provenance records containing its source occurrences. 

Each skill is assigned one primary capability category $c_i$. We denote the set of seven capability categories by $\Gamma$, which organizes the collection according to the main code artifact or development activity described by a skill. General \& Functional Code covers general implementations and function-level code. Tools, SDKs \& Generators covers utilities, software development kits, and code generators. Data \& Infrastructure covers data processing code, database operations, and executable infrastructure configuration. Testing covers test code generation. Application Components covers components within applications. Web \& API covers web functionality and API development. Code Migration covers transformations of existing code across interfaces, frameworks, or languages.

For $c\in\Gamma$, its pool is
\begin{equation}
    S_c=\{s_i\in S\mid c_i=c\}.
    \label{eq:css_pool}
\end{equation}
These pools are logical views of the shared collection, not separately duplicated skill stores. A task may require several capabilities and can therefore search the union of multiple pools. This organization permits the server to restrict matching to a relevant candidate space while retaining access to complementary capabilities.

\subsection{SKILLSERVER-CLIENT TRANSACTIONS}
\label{sec:css_transactions}

The interaction consists of three steps: FASTADAPT constructs the task request and its retrieval constraints; the SkillServer selects admissible skills; and the client receives the returned skills and decides which ones to use for code generation.

\subsubsection{FASTADAPT}
\label{sec:css_fastadapt}

FASTADAPT is a task adaptation component of CSS that uses the client model to construct a structured skill search request. Given a target coding task $q$, FASTADAPT extracts the task information and task-specific retrieval constraints required for skill search. Specifically, it produces a task query $q'$ containing the task description and available coding context, identifies the required capability categories $C$, and determines the maximum number $K$ of returned skills, the token budget $B$ for the served skill context, and the relevance threshold $\tau$.

We denote the FASTADAPT operation using client model $M$ by $\operatorname{FA}_{M}$. The search request $z$ is
\begin{equation}
    z
    =
    \operatorname{FA}_{M}(q)
    =
    (q',\theta),
    \qquad
    \theta=(r,K,B,C,\tau),
    \label{eq:css_fastadapt}
\end{equation}
where $M$ denotes the client model, and $r$ denotes the retrieval method, which is not extracted from the target task and can be configured by the user to choose among different retrieval methods.

For a task involving multiple activities, $C$ may contain multiple capability categories, while $C=\Gamma$ allows retrieval from the entire skill collection. If no capability categories are specified, CSS searches the full collection and if count, token, or relevance constraints are unspecified, CSS imposes no additional limits on the retriever's results.

\subsubsection{SKILL SELECTION}
\label{sec:css_selection}

The server receives $z$ and constructs the candidate collection

\begin{equation}
    S_C
    =
    \bigcup_{c\in C} S_c.
    \label{eq:css_candidate_pool}
\end{equation}
Because $C$ can contain multiple capability categories, $S_C$ can combine skills from several capability pools. The server then applies the configured retrieval method $r$ to match the normalized query against the prepared skill collection.

CSS treats retrieval as a pluggable method interface. Different retrieval methods can use different matching, scoring, or relation-traversal mechanisms, but they all operate on the same prepared collection and return the most relevant skills through the same serving interface. For a request, the returned skills are given by
\begin{equation}
    S_q
    =
    R(q',S_C;\theta),
    \label{eq:css_selection}
\end{equation}
where $R$ is the common CSS retrieval interface, $q'$ is the normalized query, $S_C$ is the candidate skill collection, and $\theta=(r,K,B,C,\tau)$ denotes the retrieval constraints. 

The server returns the selected skills in descending relevance order under the configured retrieval method. If fewer candidates satisfy the configured conditions, CSS may return fewer than $K$ skills. 

\subsubsection{CLIENT WITH SKILLS}
\label{sec:css_client}

After receiving $S_q$, the client combines the original task and coding context with the served skill instructions. Let $P(q)$ denote its original coding prompt and $A(S_q)$ denote the serialized skill instructions according to the order determined by the retrieval results. The augmented context is

\begin{equation}
    P_{\mathrm{CSS}}(q)
    = \operatorname{Compose}\bigl(P(q),A(S_q)\bigr),
    \qquad
    y \sim M\bigl(\cdot \mid P_{\mathrm{CSS}}(q)\bigr),
    \label{eq:css_client}
\end{equation}

where $\operatorname{Compose}$ combines the original task prompt with the
returned skill instructions, and $y$ is the program generated by the client
model $M$.

The client agent decides which served skills to use, whether individually, in combination, or not at all. CSS determines which resources are made available, while the client determines how they are used during code generation. This separation allows the same prepared collection and serving interface to support different coding agents, with the benefit of served skills evaluated by the correctness of the resulting program.

\vspace{-5pt}
\section{Experiments}
\label{sec:experiments}

We evaluate Code Skill Server (CSS) along four dimensions: its effectiveness across code generation Settings, its impact on individual tasks, its generalization across models, and the effect of the skill corpus on serving performance.

\subsection{Experimental Setup}
\label{sec:experimental_setup}

\paragraph{Benchmarks.}
We evaluate on BigCodeBench \citep{zhuo2025bigcodebench} and CoderEval-Python \citep{yu2024codereval}. The main BigCodeBench comparison uses the same tasks across all methods. We group these tasks into seven domains according to their primary dependencies: Computation, General, Visualization, System, Time, Network, and Cryptography. CoderEval-Python contains 230 code generation tasks drawn from real software projects. We group them by contextual dependency level into L1--L2, L3--L4, and L5--L6. Together, the benchmarks cover function-level generation and generation within an existing project context.

\begin{figure}[t]
    \centering
    \includegraphics[width=\linewidth]{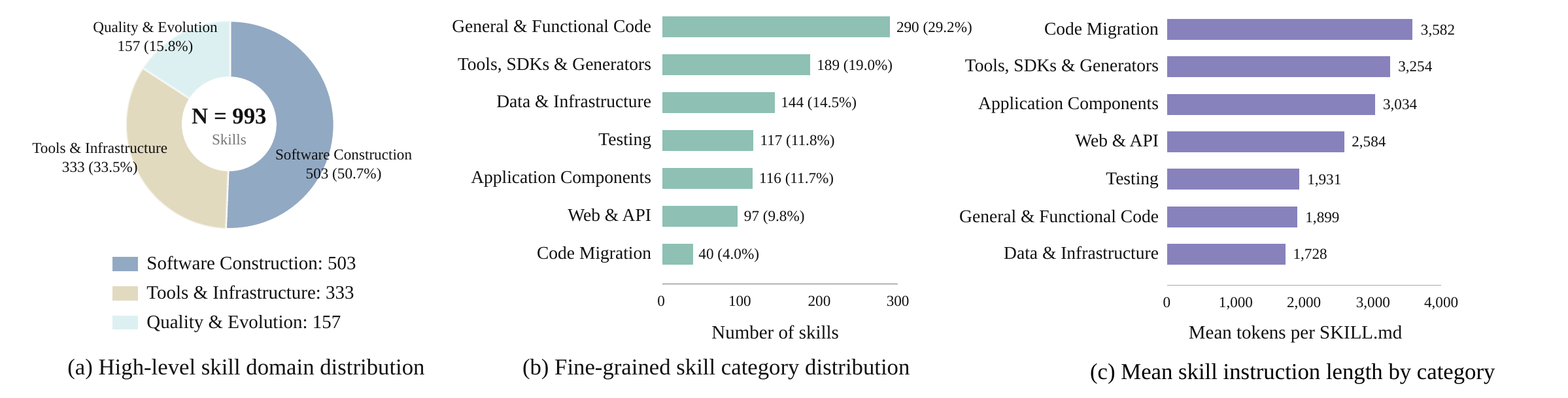}
    \caption{Distribution and textual characteristics of the 993-skill evaluation snapshot. (a) Distribution across three high-level domains. (b) Number and proportion of skills in seven fine-grained categories. (c) Mean skill instruction length by category, measured in tokens per SKILL.md file.}
    \label{fig:namber}
    \vspace{-2pt}
\end{figure}

\paragraph{Skill corpus.}
CSS collects candidate skills from public sources, including GitHub, Skills.sh, and SkillsMP. It parses and normalizes their contents, removes duplicates, and filters for code generation capabilities. The retained skills are organized into three high-level domains and seven fine-grained capability categories. For evaluation, we freeze a snapshot of 993 skills so that all retrieval methods use the same candidate corpus; this snapshot size is not an upper bound on the corpus CSS can maintain. Figure~\ref{fig:namber} shows its category distribution and mean instruction length.

\paragraph{Comparison settings.}
We compare No-Skill, BM25 \citep{robertson2009probabilistic}, Graph-of-Skills (GoS) \citep{liu2026graphofskills}, and SkillTrace \citep{yao2026skilltrace}. No-Skill provides no external skill to the coding agent, while the other settings retrieve relevant skills from the same CSS corpus and provide them through a common serving format. This setup ensures that all skill-augmented methods operate over the same underlying skill corpus and serving interface.

\paragraph{Models and evaluation.}
The main experiments use GPT-5.6 \citep{openai2026gpt56} as the primary model. For cross-model evaluation, we additionally use GLM-5.3-Flash \citep{zai2026glm53}, DeepSeek-V4-Flash \citep{deepseek2026v4flash}, MiniMax-M2.7 \citep{minimax2026m27}, and Qwen3-Coder-480B \citep{qwen2025qwen3coder}, with SkillTrace fixed as the skill-augmented setting. Following BigCodeBench \citep{zhuo2025bigcodebench}, we report Pass@1 as the primary evaluation metric. For each task, a single final solution is generated and counted as correct only if it passes all benchmark tests. All methods are evaluated on the same task inputs and with the same evaluation procedure. Reference solutions, hidden tests, and test outcomes are excluded from the retrieval query and the agent's input.

\vspace{-5pt}

\subsection{Effectiveness Across Code Generation Settings}
\label{sec:effectiveness}

CSS-served skills consistently improve code generation performance across both benchmarks. On BigCodeBench, all three skill-augmented settings outperform the No-Skill baseline in overall Pass@1, with the best result increasing from 53.42\% to 60.88\%. On CoderEval-Python, the best skill-augmented setting improves the overall pass rate from 33.04\% to 46.96\%. These results show that providing task-relevant skills can substantially improve the functional correctness of generated code across both standalone function generation and project-context code generation. The overall results are reported in Table~\ref{tab:bcb_main}.

The benefit of skills is also observed across different task categories rather than being concentrated in a small subset of tasks. On BigCodeBench, SkillTrace improves over No-Skill in all seven domains, including Computation, General, Visualization, System, Time, Network, and Cryptography. The largest improvement occurs in Network, where Pass@1 increases from 40.91\% to 62.12\%, corresponding to a gain of 21.21\%. Improvements are also observed in domains such as Computation and Time, indicating that retrieved skills can provide useful implementation guidance across diverse programming requirements. The domain-level results are shown in Table~\ref{tab:bcb_main}.

\setlength{\dbltextfloatsep}{0pt}
\begin{table}[t]
\centering
\caption{\textbf{Main results on BigCodeBench and CoderEval-Python.}
All values are Pass@1 (\%). Best results are shown in bold.}
\label{tab:bcb_main}

\small
\setlength{\tabcolsep}{3.0pt}
\renewcommand{\arraystretch}{1.15}

\resizebox{0.95\textwidth}{!}{%
\begin{tabular}{l|cccccccc|cccc}
\toprule
\multirow{2}{*}{\textbf{Method}}
& \multicolumn{8}{c|}{\textbf{BigCodeBench}}
& \multicolumn{4}{c}{\textbf{CoderEval-Python}} \\
\cmidrule(lr){2-9}
\cmidrule(lr){10-13}
& \textbf{Net.}
& \textbf{Crypto.}
& \textbf{Comp.}
& \textbf{Gen.}
& \textbf{Visual.}
& \textbf{System}
& \textbf{Time}
& \textbf{Overall}
& \textbf{L1--L2}
& \textbf{L3--L4}
& \textbf{L5--L6}
& \textbf{Overall} \\
\midrule

No-Skill
& 40.91 & 55.10 & 55.80 & 62.22
& 51.36 & 48.48 & 49.12 & 53.42
& 47.62 & 21.05 & 32.97 & 33.04 \\

BM25
& 51.52
& \cellcolor{bestgreen}\textbf{69.39}
& 57.14
& 62.96
& \cellcolor{secondgreen}56.36
& \cellcolor{bestgreen}\textbf{61.21}
& \cellcolor{secondgreen}54.39
& 58.33
& \cellcolor{secondgreen}50.79
& 35.53
& 26.37
& 36.09 \\

GoS
& \cellcolor{secondgreen}59.09
& 59.18
& \cellcolor{secondgreen}60.71
& \cellcolor{secondgreen}68.15
& \cellcolor{secondgreen}56.36
& 54.55
& \cellcolor{secondgreen}54.39
& \cellcolor{secondgreen}59.39
& \cellcolor{bestgreen}\textbf{63.49}
& \cellcolor{bestgreen}\textbf{43.42}
& \cellcolor{bestgreen}\textbf{38.46}
& \cellcolor{bestgreen}\textbf{46.96} \\

SkillTrace
& \cellcolor{bestgreen}\textbf{62.12}
& \cellcolor{secondgreen}65.31
& \cellcolor{bestgreen}\textbf{61.16}
& \cellcolor{bestgreen}\textbf{68.89}
& \cellcolor{bestgreen}\textbf{57.27}
& \cellcolor{secondgreen}58.18
& \cellcolor{bestgreen}\textbf{56.14}
& \cellcolor{bestgreen}\textbf{60.88}
& \cellcolor{secondgreen}50.79
& \cellcolor{secondgreen}38.16
& \cellcolor{secondgreen}35.16
& \cellcolor{secondgreen}40.43 \\

\bottomrule
\end{tabular}%
}
\vspace{-10pt}
\end{table}

CoderEval-Python further evaluates whether skills remain useful when code must be generated within an existing project context. Skill-augmented settings improve the overall pass rate over No-Skill, and the strongest result reaches 46.96\%, compared with 33.04\% without skills. Improvements are observed across the different contextual-dependency levels, suggesting that external skills can remain useful even when successful generation depends on information beyond the target function itself. Table~\ref{tab:bcb_main} reports the detailed results by contextual-dependency level.

The magnitude of the improvement varies across retrieval configurations and task settings, indicating that the usefulness of skills also depends on whether appropriate skills are successfully retrieved. Nevertheless, the consistent improvements over No-Skill across both benchmarks provide evidence that task-relevant skills can serve as effective external guidance for code generation.

\vspace{-7pt}
\subsection{Task-Level Impact Analysis}
\label{sec:task_impact}

Aggregate gains arise from substantially more failed tasks becoming successful than successful tasks becoming unsuccessful. We pair No-Skill and SkillTrace outcomes on the same BigCodeBench tasks and distinguish four cases: both pass, pass-to-fail, fail-to-pass, and both fail.

\begin{figure}[t]
    \centering
    \includegraphics[width=\linewidth]{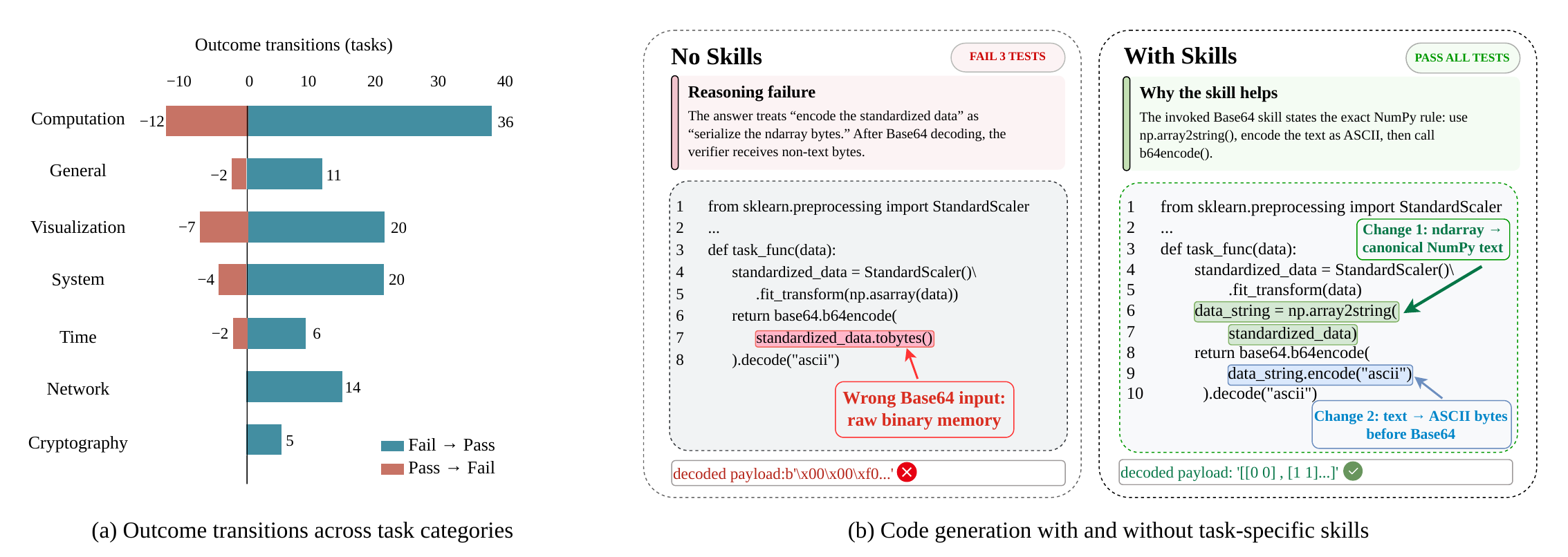}
    \caption{Impact of retrieved skills. (a) Outcome transitions across task categories. (b) Code generation with and without task-specific skills.}
    \label{fig:exm}
    \vspace{-13pt}

\end{figure}

With skills, 112 previously failed tasks become successful, while 27 previously successful tasks fail, yielding a net increase of 85 successful tasks. SkillTrace recovers 21.1\% of the tasks that fail without skills; conversely, 4.4\% of the tasks that originally pass regress. Thus, the overall improvement does not imply that every task benefits. Figure~\ref{fig:exm} (a) breaks down these paired outcomes by domain.


The transitions vary across domains. In Network, 14 tasks change from failure to success without an observed reverse transition. In Computation, 36 tasks improve while 12 regress. Skill augmentation is therefore task-dependent: providing additional skills does not guarantee a positive change on every task.

Figure~\ref{fig:exm} (b) illustrates a concrete change in the generated implementation. The task requires a standardized numeric array to be returned as Base64-encoded ASCII text. Without a skill, the generated code encodes the array's raw bytes, which do not represent the required textual form. With the relevant skill, the code first applies \texttt{np.array2string} to obtain a textual representation, encodes that text as ASCII bytes, and then applies Base64 encoding. The case shows how a skill can supply a specific procedure that changes the final implementation.

\vspace{-7pt}
\subsection{Cross-Model Generalization}
\label{sec:cross_model}

\begin{wrapfigure}{l}{0.62\textwidth}
    \centering
    \includegraphics[width=0.52\textwidth]{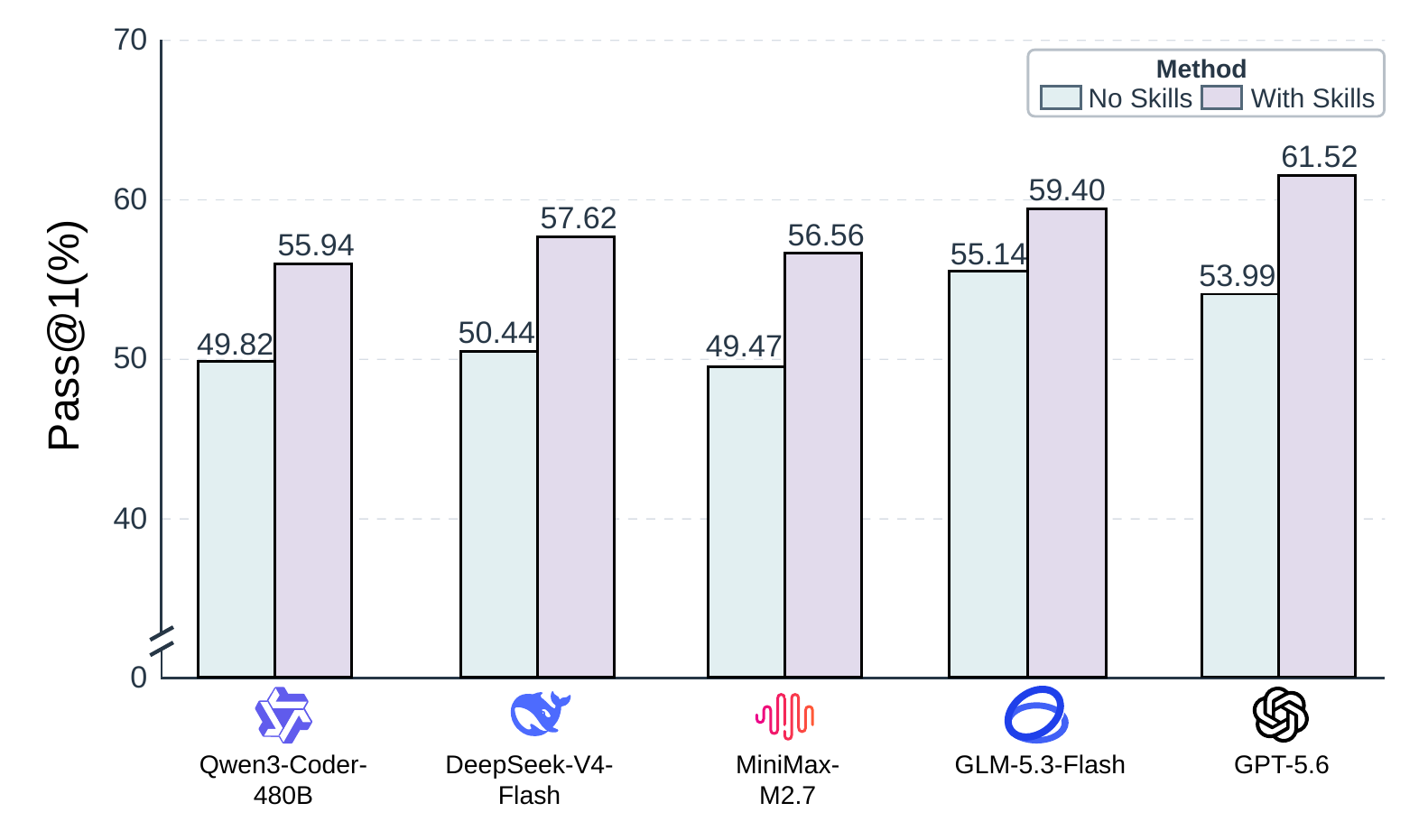}
    \vspace{-10pt}
    \caption{Pass@1 comparison between the No-Skill and With-Skill settings across five language models.}
    \label{fig:cross_model}
    \vspace{-10pt}
\end{wrapfigure}

Skill serving produces positive overall gains for all five evaluated models. To test whether the main result depends on GPT-5.6, we compare each model's No-Skill result with its result under SkillTrace. Every model improves, with gains ranging from 4.21 to 7.46 percentage points, as shown in Fig.~\ref{fig:cross_model}.

The magnitude of the gain is not a monotonic function of the No-Skill pass rate. GLM-5.3-Flash has the highest No-Skill result among the evaluated models but the smallest gain, whereas GPT-5.6 and DeepSeek-V4-Flash each gain more than seven percentage points. 

We further report category-wise Pass@1 under SkillTrace in Table~\ref{tab:cross_model_overall} to examine how the five models perform across different BigCodeBench domains. The models exhibit noticeably different performance profiles across categories. General and Cryptography are among the stronger domains for most models, whereas Time remains relatively challenging for several models.

\begin{table}[t]
\centering
\caption{\textbf{Cross-model performance comparison on BigCodeBench.}
We report the Pass@1 (\%) of SkillTrace using five models across seven task domains and overall.}
\label{tab:cross_model_overall}

\small
\setlength{\tabcolsep}{5pt}
\renewcommand{\arraystretch}{1.0}

\resizebox{0.94\textwidth}{!}{%
    \begin{tabular}{l|ccccccc|c}
        \toprule
        \modelheader
        & \textbf{Comp.}
        & \textbf{General}
        & \textbf{Visual.}
        & \textbf{System}
        & \textbf{Time}
        & \textbf{Network}
        & \textbf{Crypto.}
        & \textbf{Overall} \\
        \midrule

        \modelicon{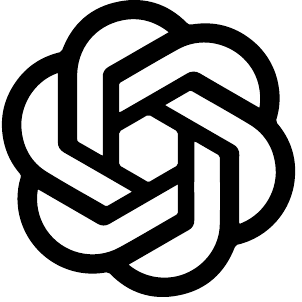}GPT-5.6
        & \cellcolor{bestgreen}\textbf{61.16}
        & \cellcolor{bestgreen}\textbf{68.89}
        & \cellcolor{bestgreen}\textbf{57.27}
        & \cellcolor{secondgreen}58.18
        & \cellcolor{bestgreen}\textbf{56.14}
        & \cellcolor{bestgreen}\textbf{62.12}
        & \cellcolor{secondgreen}65.31
        & \cellcolor{bestgreen}\textbf{60.88} \\

        \modelicon{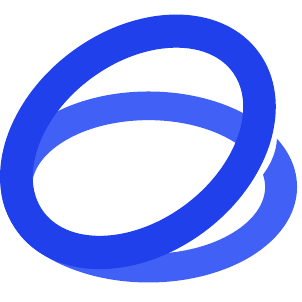}GLM-5.3-Flash
        & \cellcolor{secondgreen}58.04
        & \cellcolor{secondgreen}68.15
        & \cellcolor{secondgreen}56.82
        & 57.58
        & 45.61
        & \cellcolor{secondgreen}59.09
        & \cellcolor{bestgreen}\textbf{67.35}
        & \cellcolor{secondgreen}58.77 \\

        \modelicon{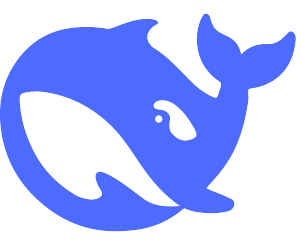}DeepSeek-V4-Flash
        & 54.24
        & 67.41
        & 53.64
        & \cellcolor{bestgreen}\textbf{60.00}
        & 50.88
        & \cellcolor{secondgreen}59.09
        & 63.27
        & 57.02 \\

        \modelicon{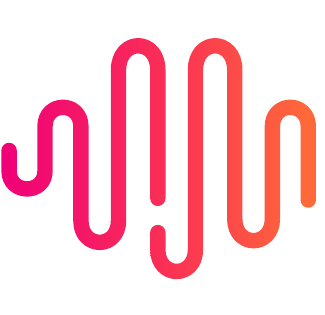}MiniMax-M2.7
        & 53.79
        & 63.70
        & 54.09
        & 56.36
        & \cellcolor{secondgreen}52.63
        & \cellcolor{secondgreen}59.09
        & 61.22
        & 55.96 \\

        \modelicon{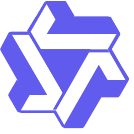}Qwen3-Coder-480B
        & 53.57
        & 62.22
        & 55.91
        & 51.52
        & 47.37
        & \cellcolor{secondgreen}59.09
        & \cellcolor{bestgreen}\textbf{67.35}
        & 55.35 \\

        \bottomrule
    \end{tabular}%
}
\vspace{-10pt}
\end{table}

\vspace{-5pt}
\subsection{Analysis of the CSS Skill Corpus}
\label{sec:corpus_analysis}

The preceding experiments establish that skill serving improves code generation across benchmarks and models. We next study how the properties of the skill corpus maintained by CSS affect downstream performance, focusing on corpus scale and corpus construction.

\begin{wraptable}[11]{r}{0.37\textwidth}
    \centering
    \caption{Effect of skill corpus size on BigCodeBench. All settings use the same 210 tasks and BM25.}
    \vspace{-6pt}
    \label{tab:skill_corpus_size}

    \small
    \setlength{\tabcolsep}{4pt}
    \renewcommand{\arraystretch}{1.12}

    \resizebox{\linewidth}{!}{%
        \begin{tabular}{lrrr}
            \toprule
            \textbf{Setting}
            & \textbf{\# Skills}
            & \textbf{Pass@1}
            & \textbf{$\Delta$} \\
            \midrule
            No-Skill  & 0   & 48.10 & -- \\
            CSS-20\%  & 199 & 51.90 & +3.81 \\
            CSS-40\%  & 397 & 56.67 & +8.57 \\
            CSS-60\%  & 596 & 57.14 & +9.05 \\
            CSS-100\% & 993 & \textbf{58.10} & +10.00 \\
            \bottomrule
        \end{tabular}%
    }
    \vspace{-10pt}
\end{wraptable}

\paragraph{Impact of corpus size.}
To isolate the effect of corpus scale while controlling evaluation cost, we use a fixed subset of 210 BigCodeBench tasks, with 30 tasks sampled from each of the seven dependency-based domains. All settings share the same task set, model, and BM25 retriever. Starting from the full 993-skill corpus, we construct nested 20\%, 40\%, and 60\% subsets by stratified sampling over the seven capability categories. As shown in Table~\ref{tab:skill_corpus_size}, Pass@1 increases from 48.10\% without skills to 51.90\%, 56.67\%, 57.14\%, and 58.10\% as the available corpus grows from 20\% to 100\%. These results indicate that broader corpus coverage increases the opportunity to retrieve task-relevant guidance.

\begin{wraptable}[8]{r}{0.37\textwidth}
    \centering
    \caption{Comparison of equal-sized skill corpora on BigCodeBench.}
    \vspace{-6pt}
    \label{tab:skill_corpus_construction}

    \small
    \setlength{\tabcolsep}{5pt}
    \renewcommand{\arraystretch}{1.12}

    \resizebox{\linewidth}{!}{%
        \begin{tabular}{lrrr}
            \toprule
            \textbf{Setting}
            & \textbf{\# Skills}
            & \textbf{Pass@1}
            & \textbf{$\Delta$} \\
            \midrule
            No-Skill & 0   & 48.10 & -- \\
            Raw-993  & 993 & 50.95 & +2.86 \\
            CSS-993  & 993 & \textbf{58.10} & +10.00 \\
            \bottomrule
        \end{tabular}%
    }
    \vspace{-10pt}
\end{wraptable}

\paragraph{Effect of corpus construction.}
We further compare two equal-sized corpora to examine whether the way the skill corpus is constructed affects downstream performance. Raw-993 is formed by randomly sampling 993 skills from the publicly collected candidate pool, whereas CSS-993 consists of 993 skills collected and organized through the CSS pipeline. Both settings use the same task set, model, BM25 retriever, and agent prompt structure. As reported in Table~\ref{tab:skill_corpus_construction}, CSS-993 achieves 58.10\% Pass@1, compared with 50.95\% for Raw-993. The result shows that the corpus constructed by CSS provides more effective support for downstream code generation than an equally sized randomly sampled corpus.

\vspace{-5pt}
\section{Conclusion} 
\label{sec:conclusion} 
In this work, we introduce Code Skill Server (CSS), a unified framework that aims to retrieve useful reusable skills for a target code generation task from a large skill corpus. In particular, CSS collects skills from public sources, converts them into a consistent representation, organizes them by capability, and exposes them through a common retrieval and serving interface. The retrieved skills are then provided to the client agent as task-specific guidance for code generation. We experimentally show that CSS is effective across different code generation settings and models, and further find that both corpus scale and corpus construction influence downstream performance. We hope that this work provides a useful foundation for scalable skill sharing and retrieval as reusable agent skills continue to grow. In the future, we aim to further improve skill selection and extend CSS to support larger and more diverse skill ecosystems.

\subsection*{AI Use Statement}

We used generative AI tools to assist with language polishing and literature search during the preparation of this manuscript. AI-assisted outputs were reviewed and verified by the authors before being incorporated into the paper. The authors take full responsibility for the final content, claims, and presentation of this work.

\subsection*{Ethics Statement}

This work does not involve human subjects or personally identifiable information. The skills used in our study are collected from publicly available sources, and the experiments are conducted solely for research purposes. We are not aware of any additional ethical concerns arising from this work.

\subsection*{Reproducibility Statement}

We provide detailed descriptions of the skill collection and organization process, retrieval settings, evaluated models, benchmarks, and evaluation protocols in the main paper. These details are intended to facilitate reproduction of our experimental results.

\bibliography{conference}
\bibliographystyle{arxiv_conference}

\end{document}